\documentclass[aps,prfluids,preprint,groupedaddress]{revtex4-2}

\usepackage{graphicx}
\usepackage{dcolumn}
\usepackage{bm}

\begin{document}


\title{Predominance of pressure transport in spatial energy budget for a mixing layer approaching absolute instability}


\author{A. B. Aadhishwaran}
\thanks{Present address: Department of Mathematics, Indian Institute of Technology Madras, Chennai 600036, India.}
\affiliation{Department of Aerospace Engineering, Indian Institute of Science, Bengaluru 560012, India.}

\author{Sourabh S. Diwan}
\thanks{Corresponding author: sdiwan@iisc.ac.in}
\affiliation{Department of Aerospace Engineering, Indian Institute of Science, Bengaluru 560012, India.}


\date{\today}

\begin{abstract}
In this letter, we report the outcome of a spatial energy budget performed for the linear convective instability of the plane incompressible mixing layer within the inviscid framework. We find that, as the critical condition for the onset of absolute instability is approached, the integrated pressure-transport term becomes increasingly more prominent as compared to the integrated production term and it dominates the energy budget completely at the critical condition. This implies that, near the threshold of absolute instability, the growth of disturbances is almost entirely due to the pressure transport mechanism (rather than the production mechanism), which is a striking result. The part of the pressure-transport term that represents the work done by the fluctuating pressure forces is seen to be primarily responsible for the observed shift in the energy balance. These results can help us better understand the physical processes causing absolute instability in a mixing layer. In particular, the redistribution of disturbance energy in streamwise direction by fluctuating pressure, which is ``non-local'' in character for incompressible flows, seems to play a key role in this respect.
\end{abstract}


\maketitle

\textit{Introduction:} The classification of instabilities in open flows into convective and absolute instability has been an important theme of research in hydrodynamic instability. The convectively unstable flows behave as noise amplifiers, whereas absolutely unstable flows exhibit an intrinsic oscillator behaviour \cite{HuerreMonke85}. This has implications for technological applications as the method by which a flow can be controlled depends on the nature of instability it undergoes.

Huerre and Rossi \cite{HuerreRossi98} have outlined the methods of determining the conditions for the onset of absolute instability, for which the dispersion relation is the essential starting point. These techniques have been applied to a variety of flows to determine the type of instability \cite{HuerreMonke90}. In particular, the plane incompressible mixing layer has been studied extensively as a prototypical example of free shear layers undergoing convective-to-absolute transition; see the references in \cite{HuerreMonke90}. Huerre and Monkewitz \cite{HuerreMonke85} were the first to determine the conditions for the onset of absolute instability in the plane mixing layer using a simple ``tanh'' model for the base velocity profile \cite{Michalke65}; see equation (\ref{eq1}).
\begin{equation} \label{eq1}
\frac{U(y)}{U_m} = 1 + \lambda \textrm{tanh}\bigg(\frac{y}{2 \delta_{\omega}}\bigg),
\end{equation}
where $U$ is the streamwise velocity, $y$ is the cross-stream distance, $\delta_{\omega}$ is the vorticity thickness, $U_m$ is the average stream velocity $(= (U_1 + U_2)/2)$, $\lambda = (U_1 - U_2) / (U_1 + U_2)$, and $U_1$ and $U_2$ are the velocities of the faster and slower streams respectively. Huerre and Monkewitz \cite{HuerreMonke85} solved the Rayleigh equation, which governs the inviscid instability of plane parallel flows \cite{HuerreRossi98}, for the mixing-layer profile in equation (\ref{eq1}) and found that the absolute instability is triggered when $\lambda$  reaches a critical value of 1.315. These results were confirmed by the experiments of Strykowski and Nigam \cite{StrykowskiNiccum91} on an axisymmetric mixing layer with the mixing-layer thickness much smaller than the jet thickness, thereby serving as a good approximation to the plane mixing layer. They clearly observed the switch in the instability character of the flow from convective to absolute (marked by an emergence of a periodic regime with a well-defined frequency) at a critical value of $\lambda=1.34$, which is quite close to $1.315$ obtained in \cite{HuerreMonke85}; see the discussion in \cite{HuerreRossi98}. The experiments of \cite{StrykowskiNiccum91} thus supported the validity of the plane parallel approximation and inviscid formulation, inherent in the Rayleigh analysis, for studying the instability of a mixing layer. There have been extensions of the incompressible mixing layer to include effects of density stratification, compressibility, visco-elasticity, confinement etc., for determining their influence on the onset of absolute instability \cite{Healey2009,Caillol2009,RayZaki2014}. In the present work, we limit ourselves to the plane homogeneous incompressible mixing layer within the inviscid parallel instability framework.

It is worth noting that, despite the above studies, the exact mechanism responsible for the onset of absolute instability in a mixing layer (as well as in other shear flows) is not entirely clear. This is partly because most of the previous studies have focused on the behaviour of \textit{eigenvalues}, e.g. the ``spatial'' branches, as the absolute instability is approached \cite{HuerreRossi98}. The corresponding behaviour of \textit{eigenfunctions} in this limit has not been investigated in sufficient detail (except for a few recent studies on separated-flow profiles \cite{Diwan2009}, \cite{Avanci_etal2019}). In particular, the spatial energy budget for the plane mixing layer approaching absolute instability has not been reported, to the best of our knowledge. In this work, we carry out such an analysis following the treatment of  Hama et al. \cite{Hama_etal80} and numerically calculate the different terms in the spatial energy equation for the mixing layer to assess their relative importance. We find that the ``pressure-transport'' term in the spatial energy balance plays an increasingly significant role in comparison to the ``energy production'' term, as the absolute instability is approached. This shows that the production mechanism, which is supposed to be the primary source of disturbance growth for the mixing layer (or any other flow for that matter), becomes progressively less active as the onset of absolute instability is reached. Interestingly, the part of the pressure-transport term that contributes to the transport of streamwise disturbance energy shows a large variation in this limit. 


\textit{Spatial Energy Budget:} The spatial energy budget for plane parallel flows within the inviscid framework, valid for the convectively unstable flows, is given by (\cite{Diwan2009})
\begin{equation} \label{eq2}
\frac{d}{dx} \int_{y_1}^{y_2} U \bigg(\frac{\overline{{u'}^2}+\overline{{v'}^2}}{2}\bigg) dy = \int_{y_1}^{y_2} (-\overline{u'v'}) \frac{dU}{dy} dy + \frac{d}{dx} \int_{y_1}^{y_2} \frac{(-\overline{u'p'})}{\rho} dy.
\end{equation}
Here $x$ is the streamwise co-ordinate, $y_1$ and $y_2$ indicate the lower and upper limits of the domain in the cross-stream direction, $\rho$ is density, and $u'$, $v'$ and $p'$ indicate fluctuations in the streamwise velocity, cross-stream velocity and pressure respectively. The overbar reresents time averaging over one period ($=2\pi/\omega$). The modal representation of these quantities is given as $(u',v',p'/\rho) = \frac{1}{2}\{[u(y),v(y),p(y)] e^{i(\alpha x - \omega t)} + c.c.\}$, where $\alpha$ is streamwise wavenumber, $\omega$ is frequency, $t$ is time and $c.c.$ indicates the complex conjugate. The term on the L.H.S. in equation (\ref{eq2}) represents the \textit{mean} transport of the disturbance kinetic energy in the streamwise direction (to be denoted as ``K.E.''). The first term on the R.H.S. in equation (\ref{eq2}) is the production of disturbance K.E. (to be denoted as ``Prod.'') and the second term on the R.H.S. is the pressure-transport term (denoted as ``P.T.''), involving the transport of disturbance pressure by the \textit{fluctuating} streamwise velocity. Note that the pressure-transport term is a distinguishing feature of the spatial energy equation and is not present in the temporal version of the energy equation \cite{Hama_etal80}. The P.T. term can be decomposed into two parts as 
\begin{equation} \label{eq3}
 \frac{d}{dx} \int_{y_1}^{y_2} \frac{(-\overline{u'p'})}{\rho} dy = \int_{y_1}^{y_2} \overline{ \frac{-p'}{\rho} \frac{\partial u'}{\partial x}} dy + \int_{y_1}^{y_2} \overline{ \frac{-u'}{\rho} \frac{\partial p'}{\partial x}} dy.
\end{equation}
The first and second terms on the R.H.S. in equation (\ref{eq3}) will be denoted as ``P.T.-I'' and ``P.T. -II'' respectively. It can be shown that P.T.-I and P.T.-II contribute  to the mean transport of the ($\overline{{v'}^2}/2$) and ($\overline{{u'}^2}/2$) respectively \cite{Diwan2009}; see equation (\ref{eq4}).
\begin{eqnarray} \label{eq4}
\frac{d}{dx} \int_{y_1}^{y_2} U \bigg(\frac{\overline{{u'}^2}}{2}\bigg) dy &=& \int_{y_1}^{y_2} (-\overline{u'v'}) \frac{dU}{dy} dy + \int_{y_1}^{y_2} \overline{ \frac{-u'}{\rho} \frac{\partial p'}{\partial x}} dy , \\ \nonumber
\frac{d}{dx} \int_{y_1}^{y_2} U \bigg(\frac{\overline{{v'}^2}}{2}\bigg) dy &=& \int_{y_1}^{y_2} \overline{ \frac{-p'}{\rho} \frac{\partial u'}{\partial x}} dy.
\end{eqnarray}
The different terms in equations (\ref{eq2}) and (\ref{eq3}) have been obtained by numerical integration after substituting the modal representations of fluctuations in the respective terms. The spatial eigenvalues ($\alpha = \alpha(\omega)$; $\omega$ real) and eigenfunctions ($u(y),v(y)$) are calculated by numerically solving the Rayleigh equation using the shooting method; the code has been validated against the known results in the literature. $p(y)$ is calculated from $u(y)$ and $v(y)$ using the linearized momentum equations \cite{Diwan2009}.


The Rayleigh analysis has been performed on the ``tanh'' profiles (equation \ref{eq1}) for four different values of $\lambda$; see figure 1(a). A pinch-point analysis is carried out and it is confirmed that this profile becomes absolutely unstable at $\lambda = 1.315$ \cite{HuerreMonke85}. The integral terms in equations (\ref{eq2}) and (\ref{eq3}) are calculated for the four values of $\lambda$ and are listed in table \ref{tab1}. The Prod. and P.T. terms are expressed as percentages of the K.E. term. As can be seen from table (\ref{tab1}), the balance, K.E. = Prod. + P.T., is satisfied to within $0.1\%$ or better, validating the Rayleigh code. As $\lambda$ increases from 0.5 to 1.314, the contribution of Prod. reduces significantly and there is a substantial gain in P.T.; both show a monotonic variation with $\lambda$. At $\lambda = 1.314$, which is on the verge of becoming absolutely unstable, P.T. dominates the energy balance entirely, with Prod. accounting for only $6\%$ of K.E. In other words, at the critical condition of absolute instability for the plane mixing layer, almost entire contribution to K.E. comes from P.T., which is a striking result. Table \ref{tab1} also includes the individual contributions of P.T.-I and P.T.-II to the pressure-transport term (as percentages of K.E.). P.T.-I shows a weak variation with increase in $\lambda$, decreasing from about $50\%$ to $38\%$ over the range of $\lambda$ considered. On the other hand, P.T.-II shows a significant increase (including a sign change) from $-40\%$ to $56\%$ as $\lambda$ increases from $0.5$ to $1.314$. Thus the increasing predominance of P.T. as the absolute instability is approached is entirely due to P.T.-II (with P.T.-I in fact having an opposite effect of decresing the magnitude of P.T. with increasing $\lambda$).
\begin{table}
\caption{Integral terms in the spatial energy budget for the mixing-layer profiles (equation \ref{eq1}) using Rayleigh analysis ; $\omega_{ma}$ is the most-amplified frequency. The percentages are expressed in terms of K.E.}
\label{tab1}       
\begin{tabular}{|c|c|c|c|c|c|c|}
\hline
$\lambda$ & $U_2/U_1$ & $\omega_{ma}$ & Prod. (\%) & P.T. (\%) & P.T.-I (\%) & P.T. -II (\%)  \\
\hline
0.5   & 0.33   & 0.2195 & 90.69 & 9.33  & 49.37 & -40.04 \\
1     &  0     & 0.209  & 59.06 & 40.97 & 44.56 & -3.59 \\
1.2   & -0.091 & 0.2005 & 34.75 & 65.26 & 40.97 & 24.29 \\ 
1.314 & -0.136 & 0.1925 &  6.01 & 94.09 & 38    & 56.09 \\
\hline
\end{tabular}
\end{table}

\begin{figure} \label{fig1} 
\centering
  \includegraphics[width=0.35\textwidth]{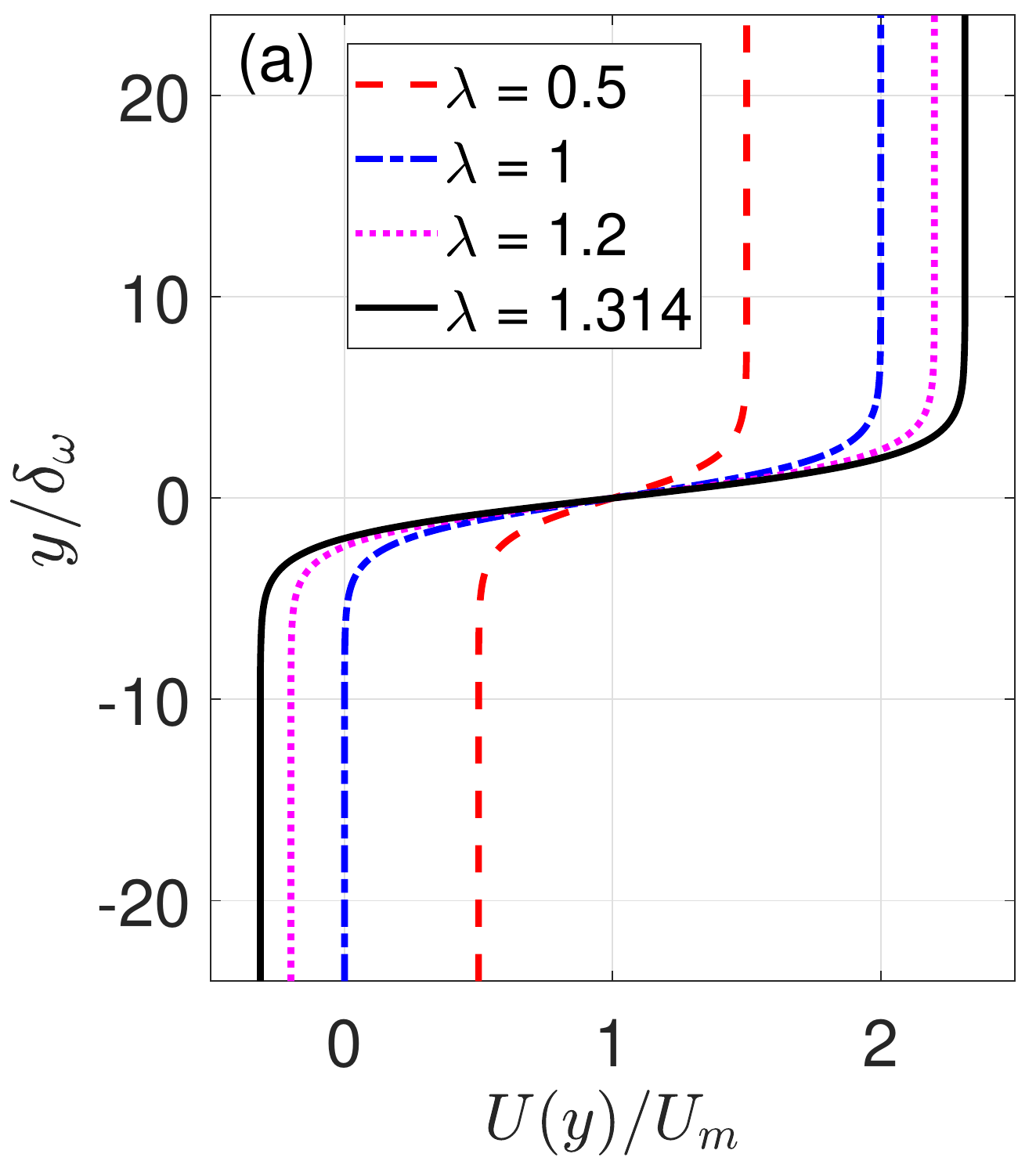}
	\\
  \includegraphics[width=0.35\textwidth]{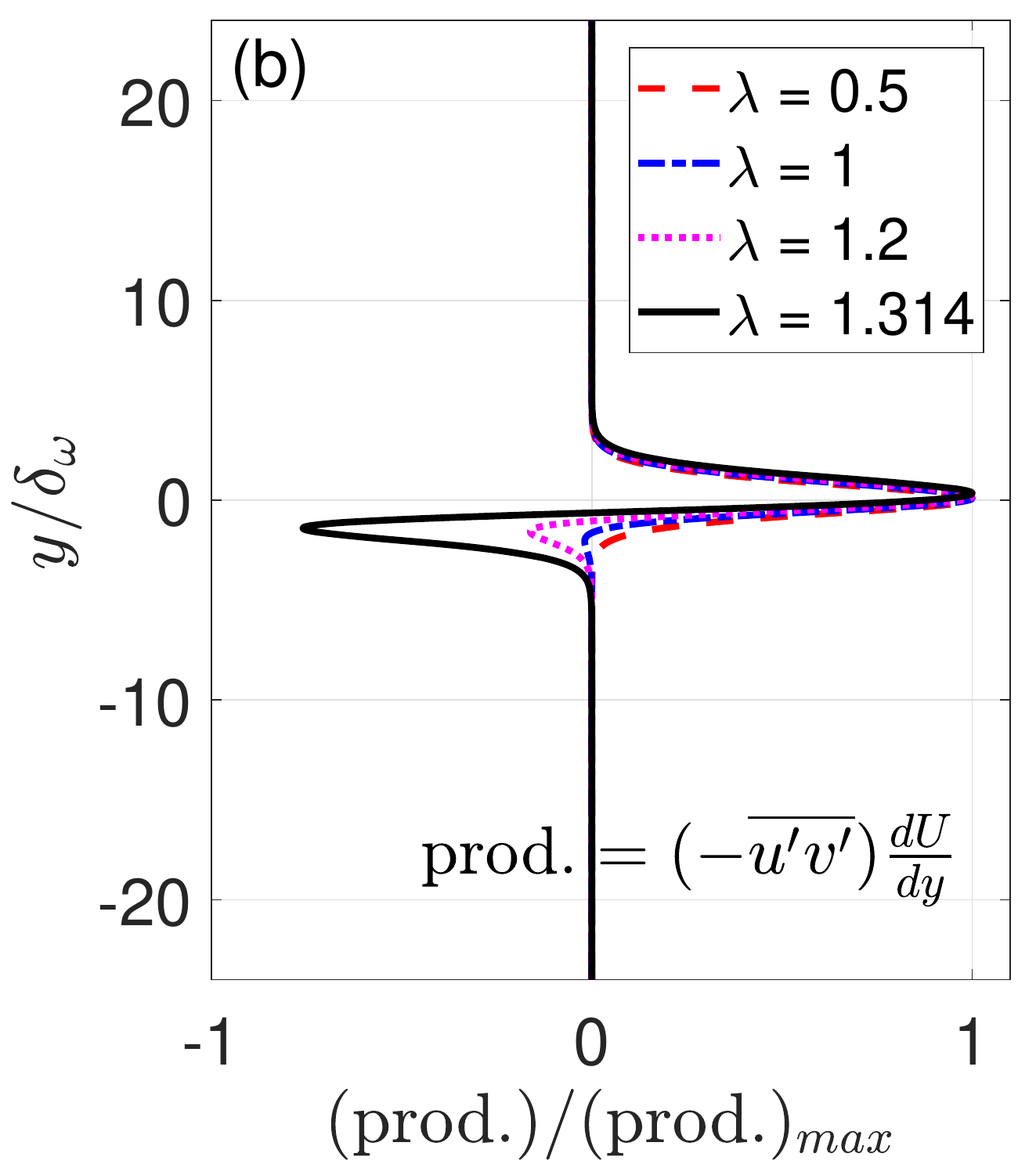}
	\includegraphics[width=0.35\textwidth]{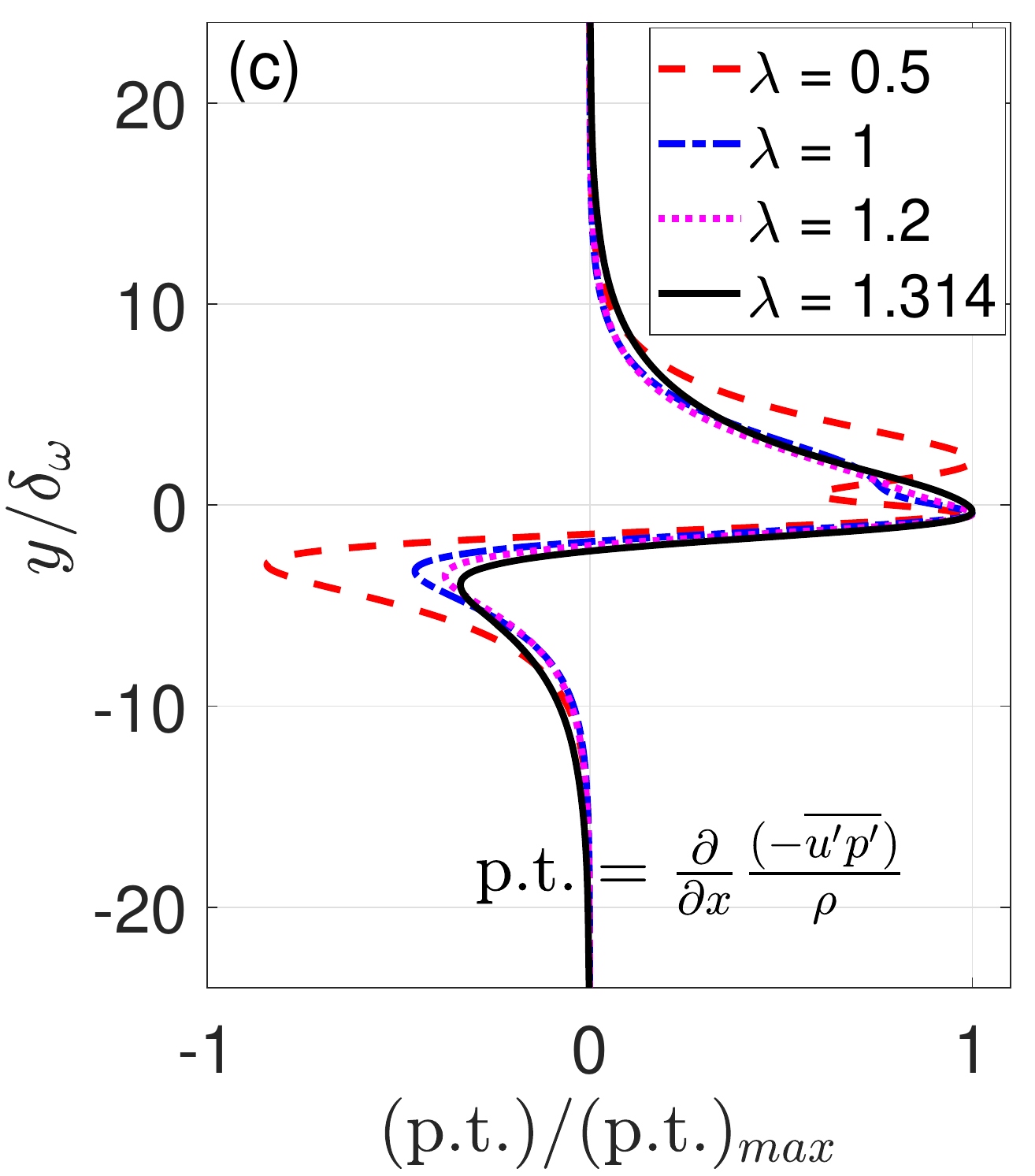}
	\includegraphics[width=0.35\textwidth]{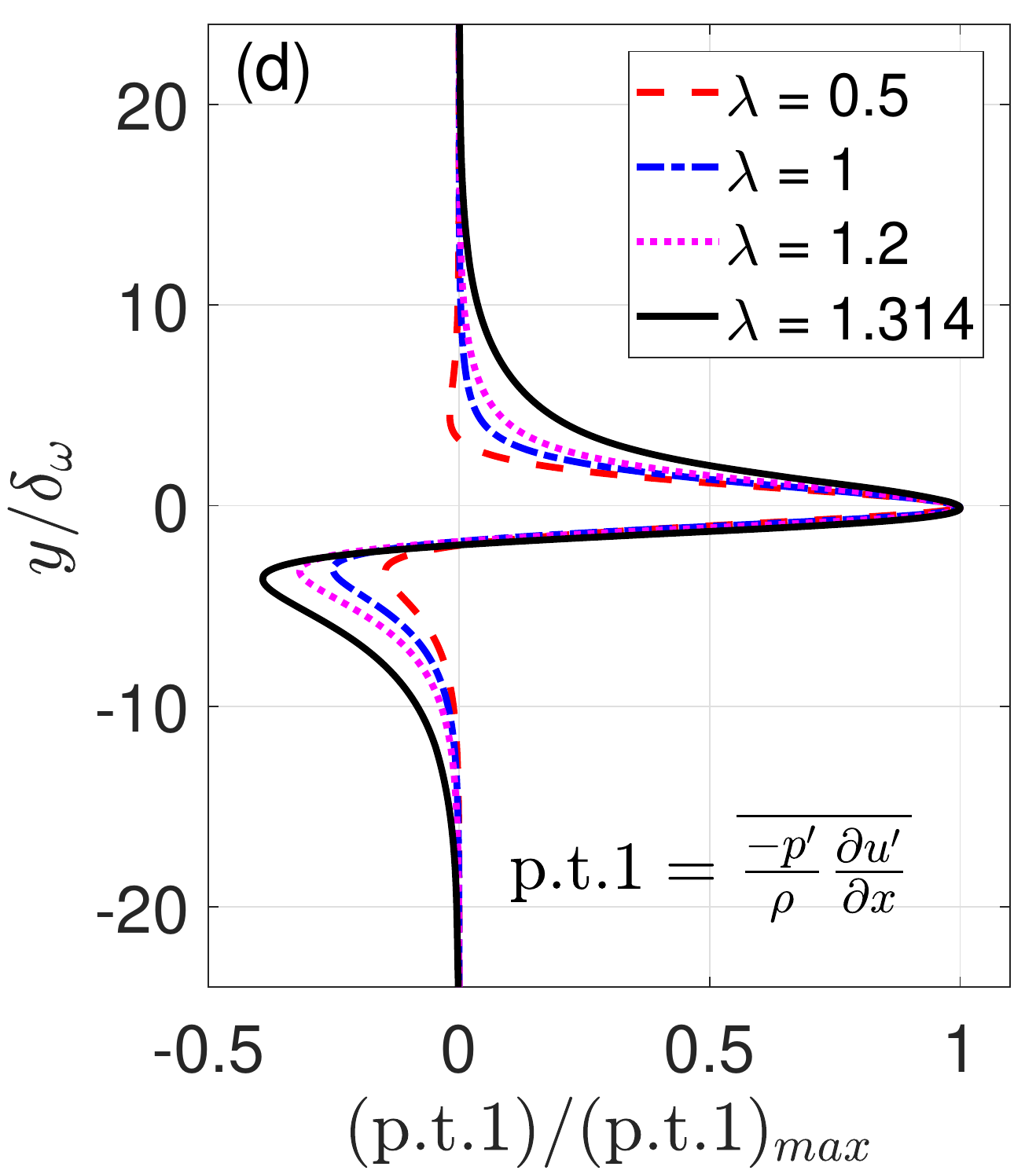}
	\includegraphics[width=0.35\textwidth]{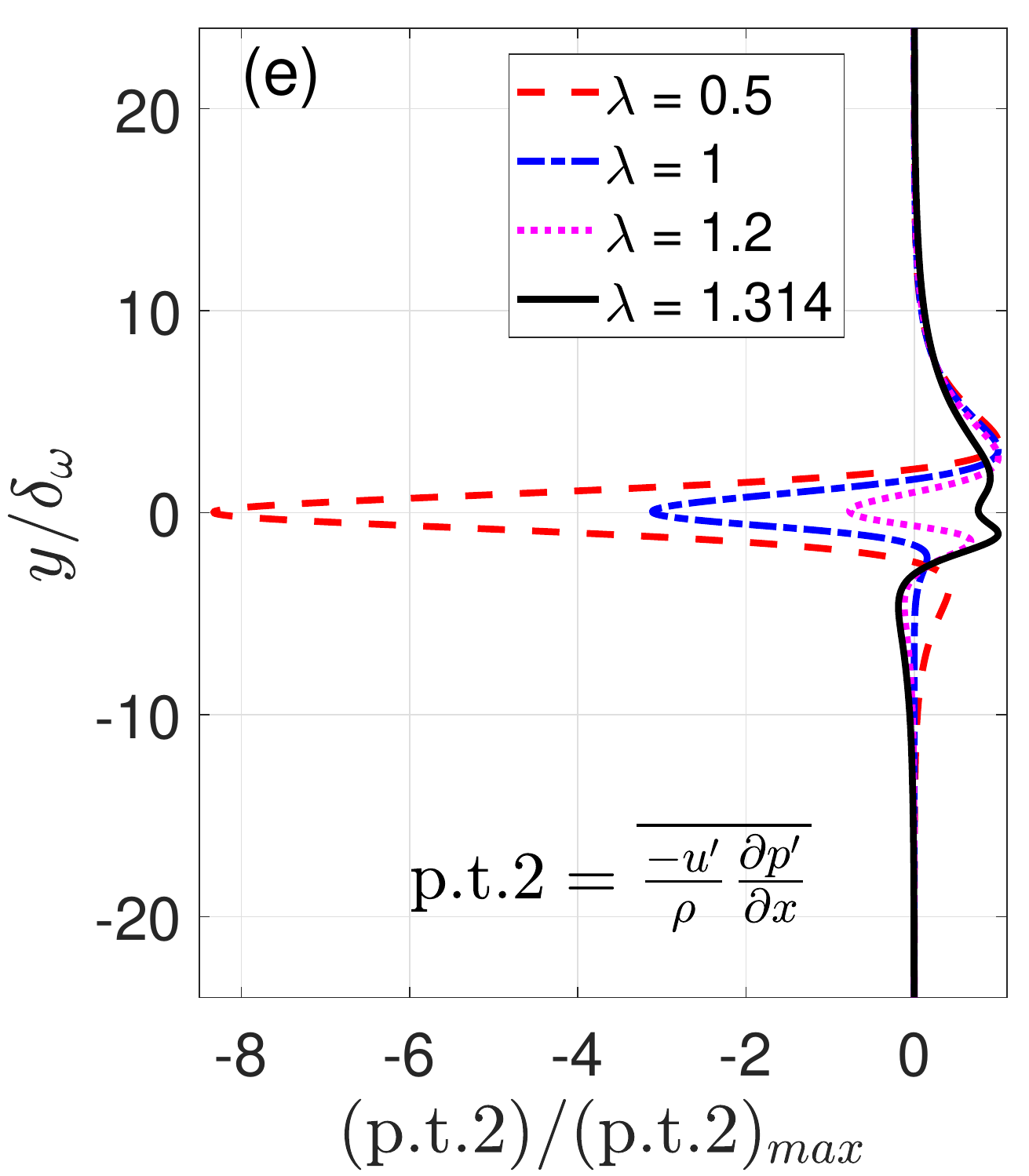}
\caption{(Color online) (a): Base-flow velocity profiles for the ``tanh'' mixing layer (equation \ref{eq1}). (b) and (c): Cross-stream profiles of the energy production and pressure-velocity correlation terms. (d) and (e): Cross-stream profiles of the constituent terms $\textrm{p.t.1}$ and $\textrm{p.t.2}$ of the pressure-velocity correlation term plotted in (c). All the quantities in (b-e) are normalized by their respective peak values.}
 \end{figure}

Figure 1(b) and 1(c) shows the distribution of the integrands in the Prod. and P.T. integrals in equation (\ref{eq2}), i.e. the profiles of energy production and pressure transport. The production profiles show a region of negative production for $\lambda > 1$, whose extent increases with increasing $\lambda$ (figure 1b). This region makes an increasingly negative contribution to the integral production as $\lambda$ approaches $1.314$, explaining the significant decrease in Prod. in this limit (table \ref{tab1}). The pressure-transport profiles exhibit regions of positive and negative values for all $\lambda$ and show a considerably larger spread in the cross-stream direction in comparison with the production profiles (figure 1b,c). The production profiles are non-zero in the interval $y/\delta_{\omega} \approx \pm 4$ (figure 2b), which is the region where most of the base-velocity shear is concentrated (figure 1a). On the other hand, the pressure-transport profiles reach out to $y/\delta_{\omega} \approx \pm (15-20)$ depending upon $\lambda$ (higher the reach larger the $\lambda$), which is 4 to 5 times larger than the reach of the production profiles. This shows that the pressure fluctuations continue to remain correlated with the streamwise velocity fluctuations even in the regions of negligible mean shear. This is an interesting observation which needs to be investigated further.

The profiles of the two constituent pressure-transport terms ($\mathrm{p.t.1}$ and $\mathrm{p.t.2}$) are shown in figure 1(d) and (e). The distribution of $\mathrm{p.t.1}$ is qualitatively similar to that of $\mathrm{p.t.}$, with the $\mathrm{p.t.1}$ profiles showing a gradual evolution as $\lambda$ increases from $0.5$ to $1.314$ (figure 1d). The $\mathrm{p.t.2}$ profiles, on the other hand, show a significant evolution with increase in $\lambda$ (figure 1e). For $\lambda=0.5$ there is a large negative region of $\mathrm{p.t.2}$ near $y=0$, flanked on either side by a positive lobe of much smaller magnitude. As $\lambda$ increases this negative region shrinks in size progressively and the relative contribution of the positive lobes increases (figure 1e). At $\lambda=1.314$, the $\mathrm{p.t.2}$ profile is dominated by positive values (in a normalized sense), with the negative region shrinking to a negligible size. There is also a considerable qualitative difference in the shape of this profile at $\lambda=1.314$ as compared to other $\lambda$ values (figure 1e). This behaviour is again consistent with the significant increase in the contribution of P.T.-II to the spatial energy balance, along with a sign change, as $\lambda$ increases (table \ref{tab1}).

It is worth noting that the terms P.T.-I and P.T.-II (or their corresponding integrands, $\mathrm{p.t.1}$ and $\mathrm{p.t.2}$) have distinct physical interpretations. P.T.-I involves the pressure-strain correlation term, which is known to be responsible for the inter-component transfer of energy as well as the observed anisotropy in turbulent shear flows \cite{TennekesLumley72}. P.T.-II represents the work done by the fluctuating pressure gradients on the streamwise velocity fluctuations. The above results (table \ref{tab1} and figure 1) indicate that the pressure-strain term does not play an important role in the onset of absolute instability in the plane mixing layer. On the other hand, the pressure-work term plays an increasingly prominent role as the absolute instability is approached. In fact, for $\lambda = 0.5$ and $1$, P.T.-II is negative implying that the work is done \textit{against} the fluctuating pressure force, whereas for $\lambda = 1.2$ and $1.314$, P.T.-II is positive meaning the work is done \textit{by} the fluctuating pressure force. Equation (\ref{eq4}) and table \ref{tab1} show that there is an approximate equi-partition of K.E. into the transport of streamwise and cross-stream components for $\lambda = 0.5$. This balance is upset as the mixing layer approaches absolute instability, with the transport of streamwise kinetic energy (equal to Prod.+P.T.-II) becoming more dominant at $\lambda = 1.314$ (table \ref{tab1}).

\textit{Concluding Remarks:} The spatial energy budget for the plane incompressible mixing layer (using ``tanh'' profiles) reveals that the integrated pressure-transport term becomes increasingly more prominent as compared to the integrated production term as the absolute instability is approached. At the threshold of absolute instability, the former completely overwhelms the latter implying that the streamwise pressure transport emerges as a new mechanism for disturbance amplification in a mixing layer. The production mechanism (believed to be the primary cause of disturbance growth in unstable flows) is much more subdued near the onset of absolute instability; its contribution to energy budget is $91\%$ at $\lambda = 0.5$, which reduces to mere $6\%$ at $\lambda = 1.314$. The cross-stream distribution of the pressure-transport term is found to be more spread out compared to the more compact distribution of the production term. The part of the pressure-transport term, $\overline{ \frac{-u'}{\rho} \frac{\partial p'}{\partial x}}$, which represents the work done by the fluctuating pressure forces (and which contributes to the mean transport of \textit{streamwise} disturbance kinetic energy), is seen to be primarily responsible for the observed shift in the energy balance. The other part, $\overline{ \frac{-p'}{\rho} \frac{\partial u'}{\partial x}}$, which is the pressure-strain correlation term, plays a relatively unimportant role as the absolute instability is approached. 

These are interesting results which could help us identify the physical processes that cause the transition from convective to absolute instability in the plane mixing layer. In particular, the dynamics of fluctuating pressure and its gradient, and their correlation with the fluctuating velocity seem to play a key role in this transition. This is conceivable as, for incompressible flows, pressure is a ``non-local'' quantity \cite{Kim89} and therefore, for the incompressible mixing layer studied here, the velocity fluctuations coupling with those of pressure can trigger absolute instability. This aspect has received little attention in the literature and needs further investigation. It is also of interest to extend the present analysis to other flows undergoing absolute instability, both incompressible and compressible. This exercise is currently underway.

\begin{acknowledgements}
We thank Prof O. N. Ramesh, Department of Aerospace engineering, Indian Institute of Science, Bengaluru for suggesting the problem and useful discussions. SSD acknowledges financial support from IISc, Bengaluru in the form of a start-up grant (No. 1205010620). 
\end{acknowledgements}

%
%

\bibliography{refs_PRF_tanh}   


\end{document}